\documentclass[%
 amsmath,amssymb,
 aip,
]{revtex4-2}

\usepackage{graphicx}
\usepackage{dcolumn}
\usepackage{bm}
\usepackage{algorithm}
\usepackage{algpseudocode}
\usepackage{xcolor}

\usepackage{graphicx}
\usepackage{subcaption}

\begin{document}

\title{Accurate wetting dynamics via a conservative Allen-Cahn based lattice Boltzmann approach for multiphase flows}

\newcommand{\RomaTreAff}{Department of Civil, Computer Science and Aeronautical Technologies Engineering,
Roma Tre University, Via Vito Volterra, 00146 Rome, Italy}

\newcommand{\CNRAff}{Istituto per le Applicazioni del Calcolo,
Consiglio Nazionale delle Ricerche, Via dei Taurini 19, 00185 Rome, Italy}

\author{Paolo Bello}
\affiliation{\RomaTreAff}

\author{Luca Mander}
\affiliation{\RomaTreAff}

\author{Marco Lauricella}
\affiliation{\CNRAff}

\author{Gianmarco Guglielmo}
\affiliation{\CNRAff}

\author{Michele La Rocca}
\affiliation{\RomaTreAff}

\author{Andrea Montessori}
\email{andrea.montessori@uniroma3.it}
\affiliation{\RomaTreAff}

\date{\today}

\begin{abstract}
In this work we propose a local geometric wetting boundary condition for a conservative Allen--Cahn-based lattice Boltzmann framework. The prescribed contact angle is imposed through ghost phase-field values constructed from a locally reconstructed wall normal and a donor-fluid extrapolation. The ghost-node wetting update is local, geometrically consistent, and
compatible with thread-safe large-scale implementations, while phase-field
mass is controlled through a separate global volume correction. Validation includes static contact-angle tests, short-time droplet spreading, impact on a hydrophobic surface, and gravity-driven motion through a sharp-edged orifice. The simulations recover the imposed equilibrium angles, reproduce contact-angle-dependent spreading exponents between approximately \(1/2\) and \(1/4\), and follow the classical \(We^{1/4}\) maximum-deformation scaling. The model also captures the transition between capture, release, and release with breakup. The proposed approach provides an accurate and scalable framework for wetting-controlled flows in complex geometries.
\end{abstract}

\maketitle


\section{Introduction}

Wetting is one of the key mechanisms through which solid boundaries control the dynamics of multiphase flows \cite{bonn2009wetting,gennes2004capillarity,dussan1979spreading}. The affinity between a fluid interface and a solid surface governs the pathways by which droplets spread or retract, liquids invade or bypass pore throats, and menisci remain pinned, depin, fragment, or pass through geometric constrictions \cite{blunt2017multiphase}. These processes are central to a broad range of natural and technological systems, including coating and printing \cite{lohse2022fundamental}, microfluidics \cite{nan2024development,zhao2016wettability}, emulsification \cite{eggersdorfer2018wetting}, enhanced oil recovery \cite{zhao2010pore,morrow1990wettability}, groundwater remediation \cite{yang2021effect}, geological carbon storage, and reactive transport in porous and fractured media \cite{ladd2021reactive}. In all these configurations, the macroscopic flow response is strongly affected by contact-line motion and by the local balance between capillary, viscous, inertial, and gravitational forces at solid boundaries.

A predictive numerical description of wetting-controlled multiphase flows is therefore challenging for at least three reasons. First, the contact angle must be imposed accurately at the diffuse-interface level while recovering the desired macroscopic wetting behavior. Second, the method must remain reliable when interfaces undergo large deformations, interact with sharp edges, or move across complex solid geometries. Third, the wetting formulation should be compatible with conservative interface transport, variable material properties, and massively parallel implementations, since realistic three-dimensional applications often involve large domains, highly irregular solid boundaries, and long-time interface evolution.

Diffuse-interface and phase-field methods provide a natural framework for this class of problems, since the interface is represented by a continuous order parameter, which takes approximately constant values in the bulk phases and varies smoothly across a thin interfacial region. This description allows topological changes, interface deformation, and capillary forces to be treated without explicit interface reconstruction \cite{anderson1998diffuse}. In particular, conservative Allen--Cahn (CAC) formulations have attracted increasing attention because they retain the interface-capturing capability of diffuse-interface methods while improving mass conservation and controlling the interface thickness through a second-order spatial operator \cite{sun2007sharp,hwang2026review}.

When coupled to lattice Boltzmann methods, CAC formulations provide a flexible
and efficient framework for simulating multiphase flows with complex boundaries,
local forcing, and large density or viscosity contrasts
\cite{lauricella2025thread,lauricella2025acclb,dinesh2019phase,
liang2018phase,geier2015conservative}, while retaining a lower computational
complexity than formulations based on the fourth-order Cahn--Hilliard equation
\cite{lee2016comparison,wang2016comparative}.

However, imposing wettability in CAC formulations remains non-trivial because
the second-order phase-field equation requires a single boundary condition for
the order parameter \cite{huang2022implementing,brown2024mass}. At an
impermeable wall, conservation requires the phase-field flux to satisfy a
no-flux condition, whereas a prescribed contact angle is commonly enforced
through an inhomogeneous Neumann condition derived either from a wall-energy
formulation or from the local interface geometry
\cite{ding2007wetting,lee2011accurate,huang2022implementing}. Except for neutral
wetting, these two conditions are not generally equivalent. Consequently, although
the CAC equation remains conservative in the bulk, the wall treatment may introduce
a small gain or loss of phase-field mass when the available boundary condition is
used to impose wettability. An additional compatibility mechanism, such as a source term, a Lagrange
multiplier, or a conservative volume-redistribution procedure, may therefore be
required \cite{huang2022implementing,brown2024mass}.

A second challenge concerns the geometrical realization of wetting boundary
conditions beyond simple planar walls. Schemes formulated for a prescribed
Cartesian wall direction may become ambiguous on voxelized solids and near
corners or sharp edges, where the local wall orientation must be reconstructed
from the surrounding fluid--solid geometry. Existing approaches address this
problem either by reconstructing ghost or virtual phase-field values from the
local interface profile \cite{huang2022simplified}, or by imposing the
order-parameter gradient directly at fluid boundary nodes through compact local
stencils \cite{zhang2023simplified}. A robust wetting formulation should
therefore combine compatibility with phase-field mass conservation, a local reconstruction of the wall geometry, and an implementation suitable for large-scale lattice
Boltzmann simulations.

In this work, building on geometric contact-angle formulations for
diffuse-interface methods \cite{ding2007wetting,huang2015wetting} and
ghost-value reconstruction strategies developed for phase-field lattice
Boltzmann models \cite{huang2022simplified}, we develop and validate a local
geometric wetting boundary condition for the conservative Allen--Cahn-based
lattice Boltzmann framework introduced in Ref.~\cite{lauricella2025thread}.
The local wall normal is reconstructed from the fluid--solid occupancy, a donor
fluid node is selected according to its alignment with this normal, and the
phase field is extrapolated along the corresponding donor-to-solid lattice
direction so that the prescribed contact angle is satisfied. In contrast to ghost-value reconstructions based on the local hyperbolic-tangent interface profile \cite{huang2022implementing}, the proposed construction relies directly on the local voxelized geometry. Compared with direct-gradient schemes \cite{zhang2023simplified}, it retains a local and explicit ghost-node formulation, while phase-field mass is controlled through a separate global constraint whose correction is redistributed over the diffuse-interface region.

The accuracy and robustness of the method are assessed through four
complementary benchmarks of increasing complexity. Static droplets on a flat
wall test the recovery of the prescribed equilibrium contact angle over a broad
range of wettabilities. Short-time spreading then examines the
contact-angle-dependent inertial--capillary dynamics, while droplet impact on a
hydrophobic surface tests the wetting treatment under finite-velocity impact
and large deformation over a range of Weber numbers. Finally, gravity-driven
droplet motion through a sharp-edged orifice is compared with experimental
regime maps. This last configuration provides a stringent test of the combined
effects of wetting, contact-line motion, edge pinning, confinement, large
deformation, and breakup.

\section{Methods}  \label{sec:methods}

\subsection{Coupled recursive-regularized thread-safe lattice Boltzmann and finite-difference MUSCL scheme for the Navier--Stokes--conservative Allen--Cahn equations}
\label{sec:tslb}

The hydrodynamics of the two-component system is solved using a thread-safe lattice Boltzmann (LB) method implemented in \textit{accLB} \cite{lauricella2025acclb,montessori2026breakdown}, coupled to an interface-capturing phase-field solver. The numerical framework combines a variable-density, variable-viscosity LB formulation for the Navier--Stokes equations with a conservative Allen--Cahn equation for the evolution of the phase field. The LB solver is designed to satisfy two main requirements. First, the update remains fully local on shared-memory architectures, such as GPUs, thereby avoiding race conditions during the streaming--collision step. Second, numerical stability at low viscosities is improved through a third-order recursive regularization of the non-equilibrium moments \cite{montessori2024high}.

\textit{Hydrodynamic limit.}
In the low-Mach- and small-Knudsen-number limits, the discrete kinetic formulation recovers the variable-density, variable-viscosity Navier--Stokes equations \cite{fakhari2017improved,zu2013phase}
\begin{align}
\partial_t \rho + \nabla\cdot(\rho \mathbf{u}) &= 0, \label{eq:conti}\\
\partial_t(\rho \mathbf{u}) + \nabla\cdot(\rho \mathbf{u}\otimes \mathbf{u}) &=
-\nabla p
+
\nabla\cdot\!\left[
\rho \nu
\left(
\nabla\mathbf{u}+(\nabla\mathbf{u})^{T}
\right)
\right]
+
\mathbf{F}_{\sigma}
+
\mathbf{F}_{\rm ext}.
\label{eq:mom}
\end{align}
Here $\rho$ denotes the mixture density, $\mathbf{u}$ the velocity, $\nu$ the kinematic viscosity, and $p$ the hydrodynamic pressure. The term $\mathbf{F}_{\sigma}$ is the capillary force density associated with the diffuse interface, while $\mathbf{F}_{\rm ext}$ accounts for possible external body forces.

\textit{Discrete kinetic equation and thread-safe update.}
We use a $q$-speed lattice stencil, specifically D3Q27 in the present implementation, with discrete velocities $\{\mathbf{c}_i\}_{i=0}^{q-1}$, quadrature weights $\{w_i\}$, and lattice time step $\Delta t=1$. The LB evolution equation reads \cite{kruger2017lattice}
\begin{equation}
f_i(\mathbf{x}+\mathbf{c}_i, t+1) =
f_i(\mathbf{x}, t)
+
\omega
\left[
f_i^{\rm eq}(\mathbf{x},t)-f_i(\mathbf{x},t)
\right]
+
S_i(\mathbf{x},t),
\label{eq:lbe}
\end{equation}
where $f_i$ are the discrete populations, $\omega$ is the relaxation frequency, and $S_i$ is the discrete forcing term \cite{guo2002discrete}. The kinematic viscosity is related to $\omega$ by
\begin{equation}
\nu =
c_s^2
\left(
\frac{1}{\omega}-\frac{1}{2}
\right),
\label{eq:nuomega}
\end{equation}
where $c_s$ is the lattice sound speed \cite{succi2018lattice}.

Following the thread-safe formulation \cite{lauricella2025thread,montessori2023thread}, both the equilibrium and non-equilibrium components of the distribution function are reconstructed from local macroscopic quantities through the Hermite expansions detailed below. Therefore, each lattice node can be updated independently, without concurrent read/write access to neighboring populations. This single-pass streaming--collision strategy avoids memory race conditions and is well suited to shared-memory parallel architectures, including GPU accelerators.

\textit{Macroscopic fields.}
We adopt the incompressible-pressure formulation \cite{zu2013phase}, in which the zeroth kinetic moment provides a dimensionless pressure-like variable $p^{*}$. The physical pressure is defined as
\begin{equation}
p =
\rho c_s^2 p^{*}.
\label{eq:pstar}
\end{equation}
The macroscopic fields are recovered as
\begin{align}
p^{*}(\mathbf{x},t)
&=
\sum_i f_i(\mathbf{x},t),
\label{eq:pstar_moment}\\
\mathbf{u}(\mathbf{x},t)
&=
\sum_i f_i(\mathbf{x},t)\mathbf{c}_i
+
\frac{1}{2}
\sum_i S_i(\mathbf{x},t)\mathbf{c}_i .
\label{eq:u_moment}
\end{align}
The density $\rho$ and viscosity $\nu$ are spatially varying quantities determined by the phase field through mixture laws.

\textit{Mixture properties and relaxation frequency.}
The local material properties are prescribed as functions of the phase field $\phi\in[0,1]$, with $\phi=0$ denoting the gas phase and $\phi=1$ the liquid phase. The density is interpolated as
\begin{equation}
\rho(\phi)
=
\rho_g
+
\left(
\rho_\ell-\rho_g
\right)\phi ,
\label{eq:rho_mix_linear}
\end{equation}
while the dynamic viscosity is interpolated linearly as \cite{fakhari2017improved}
\begin{equation}
\mu(\phi)
=
\mu_g
+
\left(
\mu_\ell-\mu_g
\right)\phi .
\label{eq:mu_mix_linear}
\end{equation}
The kinematic viscosity follows from $\nu(\phi)=\mu(\phi)/\rho(\phi)$.
The relaxation frequency is therefore local and reads
\begin{equation}
\omega(\phi)
=
\left(
\frac{1}{2}
+
\frac{\nu(\phi)}{c_s^2}
\right)^{-1},
\qquad
\tau(\phi)=\omega(\phi)^{-1}.
\label{eq:omega_phi}
\end{equation}
The same density field is used consistently in the pressure relation $p=\rho c_s^2 p^{*}$.

\textit{Third-order equilibrium.}
The equilibrium populations are constructed from a third-order Hermite expansion in Mach number \cite{montessori2024high,montessori2015lattice,malaspinas2015increasing}:
\begin{equation}
f_i^{\rm eq}
=
w_i
\Bigg[
p^{*}
+
\frac{\mathbf{c}_i\cdot\mathbf{u}}{c_s^2}
+
\frac{
\mathcal{H}_i^{(2)}
:
\left(
\mathbf{u}\otimes\mathbf{u}
\right)
}{2c_s^4}
+
\frac{
\mathcal{H}_i^{(3)}
\vdots
\left(
\mathbf{u}\otimes\mathbf{u}\otimes\mathbf{u}
\right)
}{6c_s^6}
\Bigg],
\label{eq:feq3}
\end{equation}
where ``$:$'' denotes double contraction between second-order tensors and ``$\vdots$'' denotes triple contraction between third-order tensors. The second- and third-order Hermite tensors are
\begin{align}
\mathcal{H}_i^{(2)}
&=
\mathbf{c}_i\otimes\mathbf{c}_i
-
c_s^2\mathbf{I},
\label{eq:H2}\\
\mathcal{H}_i^{(3)}
&=
\mathbf{c}_i\otimes\mathbf{c}_i\otimes\mathbf{c}_i
-
c_s^2
\mathcal{S}
\left(
\mathbf{c}_i\otimes\mathbf{I}
\right),
\label{eq:H3}
\end{align}
with $\mathbf{I}$ the identity tensor and $\mathcal{S}(\cdot)$  denoting the unnormalized sum over the three distinct index permutations.

\textit{Recursive regularization of non-equilibrium populations.}
The populations are decomposed as
\[
f_i=f_i^{\rm eq}+f_i^{\rm neq},
\]
and the non-equilibrium contribution is reconstructed up to third order as
\begin{equation}
f_i^{\rm neq}
=
w_i
\Bigg[
\frac{
\mathcal{H}_i^{(2)}
:
\mathbf{A}^{(2)}_{\rm neq}
}{2c_s^4}
+
\frac{
\mathcal{H}_i^{(3)}
\vdots
\mathbf{A}^{(3)}_{\rm neq}
}{6c_s^6}
\Bigg],
\label{eq:fneq3}
\end{equation}
where $\mathbf{A}^{(2)}_{\rm neq}$ is the symmetric second-order non-equilibrium moment, while $\mathbf{A}^{(3)}_{\rm neq}$ is obtained by Hermite recursivity \cite{szeg1939orthogonal}. At the Navier--Stokes level, the second-order non-equilibrium moment can be expressed in gradient form as
\begin{equation}
\mathbf{A}^{(2)}_{\rm neq}
=
-\frac{c_s^2}{\omega}
\left[
\nabla(\mathbf{u})
+
\nabla(\mathbf{u})^{T}
\right],
\label{eq:A2grad}
\end{equation}
and the third-order tensor is reconstructed recursively as
\begin{equation}
\mathbf{A}^{(3)}_{\rm neq}
=
\mathcal{S}
\left(
\mathbf{u}
\otimes
\mathbf{A}^{(2)}_{\rm neq}
\right).
\label{eq:A3rec}
\end{equation}
The regularization can thus be interpreted as a projection of $f_i^{\rm neq}$ onto the Hermite subspace spanned by moments up to third order. This filters higher-order ghost modes and improves robustness in low-viscosity and high-Reynolds-number regimes.

\textit{Forcing scheme and multiphase force decomposition.}
Body forces are incorporated through the Guo forcing scheme \cite{guo2002discrete},
\begin{equation}
S_i
=
w_i
\left[
\frac{\mathbf{c}_i-\mathbf{u}}{c_s^2}
+
\frac{
(\mathbf{c}_i\cdot\mathbf{u})
}{c_s^4}
\mathbf{c}_i
\right]
\cdot
\mathbf{F},
\label{eq:guo}
\end{equation}
and are implemented consistently with trapezoidal time integration by shifting the equilibrium populations, equivalently using $f_i^{\rm eq}\leftarrow f_i^{\rm eq}-\tfrac{1}{2}S_i$ in the collision step.

For multiphase flows with strong density and viscosity variations, the total force is decomposed as \cite{zu2013phase}
\begin{equation}
\mathbf{F}
=
\mathbf{F}_{\sigma}
+
\mathbf{F}_{p}
+
\mathbf{F}_{\nu}
+
\mathbf{F}_{\rm ext}.
\label{eq:Fsplit}
\end{equation}
The capillary force is written in chemical-potential form \cite{jacqmin1999calculation},
\begin{equation}
\mathbf{F}_{\sigma}
=
\mu_{\phi}\nabla\phi,
\label{eq:Fcap}
\end{equation}
where $\mu_{\phi}$ is the chemical potential associated with the phase field and defined as, $\mu_\phi=4\beta(\phi-\phi_l)(\phi-\phi_g)(\phi-\phi_0) - \kappa_\phi \partial_\alpha^2 \phi$, where $\beta=12\sigma/\delta$ and $\kappa_\phi=3\sigma\delta/2$ \cite{jacqmin1999calculation}.

Since the pressure is defined as $p=\rho c_s^2p^{*}$, its gradient can be decomposed as
\[
\nabla p
=
\rho c_s^2\nabla p^{*}
+
p^{*}c_s^2\nabla\rho .
\]
The first contribution is embedded in the equilibrium populations, whereas the second is introduced explicitly through the correction force
\begin{equation}
\mathbf{F}_{p}
=
-p^{*}c_s^2\nabla\rho .
\label{eq:Fp}
\end{equation}
To recover the variable-coefficient viscous operator in Eq.~\eqref{eq:mom}, an additional correction proportional to $\nabla\rho$ is included by exploiting the relation between the second-order kinetic moment and the deviatoric stress \cite{kruger2009shear}. In compact tensor form, this correction reads
\begin{equation}
\mathbf{F}_{\nu}
=
-\frac{\nu\omega}{c_s^2}
\left[
\sum_i
\left(
f_i-f_i^{\rm eq}
\right)
\left(
\mathbf{c}_i\otimes\mathbf{c}_i
\right)
\right]
\cdot
\nabla\rho ,
\label{eq:Fnu}
\end{equation}
where ``$\cdot$'' denotes contraction of the second-order tensor with the vector $\nabla\rho$.

\textit{Discrete derivatives.}
Gradients and Laplacians of scalar fields, such as $\rho$, $\phi$, and $\mu_{\phi}$, are evaluated using isotropic lattice-consistent finite-difference stencils \cite{thampi2013isotropic,shan2006analysis}:
\begin{align}
\nabla \Psi(\mathbf{x})
&=
\frac{1}{c_s^2}
\sum_i
w_i
\Psi(\mathbf{x}+\mathbf{c}_i)
\mathbf{c}_i,
\label{eq:grad_iso}\\
\nabla^2 \Psi(\mathbf{x})
&=
\frac{2}{c_s^2}
\left(
\sum_{i\neq 0}
w_i
\Psi(\mathbf{x}+\mathbf{c}_i)
-
(1-w_0)
\Psi(\mathbf{x})
\right),
\label{eq:lap_iso}
\end{align}
where $\Psi$ denotes a generic scalar field.

\subsubsection{MUSCL scheme discretization for the conservative Allen--Cahn equation}

\textit{Conservative Allen--Cahn equation.}
The interface is represented by a phase field $\phi\in[0,1]$, with $\phi=1$ corresponding to the liquid phase and $\phi=0$ to the gas phase. The diffuse interface is conventionally located at $\phi_0=1/2$. Following our previous work \cite{lauricella2025thread}, the evolution of $\phi$ is governed by the CAC equation \cite{chiu2011conservative}:
\begin{equation}
\frac{\partial \phi}{\partial t}
+
\mathbf{u}\cdot\nabla\phi
=
D\nabla^2\phi
-
\kappa
\nabla\cdot
\left[
\phi(1-\phi)\mathbf{n}
\right],
\label{eq:cace}
\end{equation}
where $D$ is a constant interface diffusivity and $\mathbf{n}$ is the interface normal,
\begin{equation}
\mathbf{n}
=
\frac{\nabla\phi}{|\nabla\phi|}.
\label{eq:normal}
\end{equation}
The last term in Eq.~\eqref{eq:cace} is a compressive, anti-diffusive flux that counteracts numerical diffusion and maintains a smooth but sharp hyperbolic-tangent-like interface profile across a prescribed thickness $\delta$. The compression strength is related to the diffusivity through
\begin{equation}
\kappa
=
\frac{4D}{\delta},
\label{eq:kappa_delta}
\end{equation}
so that the interface remains close to the target thickness during the simulation.

\textit{Conservative flux form.}
For numerical robustness and improved conservation at the discrete level, the advective contribution is written in conservative form. Introducing the advective flux
\[
\mathbf{F}_a=\mathbf{u}\phi,
\]
Eq.~\eqref{eq:cace} is advanced in the form
\begin{equation}
\frac{\partial \phi}{\partial t}
+
\nabla\cdot(\mathbf{u}\phi)
=
D\nabla^2\phi
-
\kappa
\nabla\cdot
\left[
\phi(1-\phi)\mathbf{n}
\right],
\label{eq:cace_conservative}
\end{equation}
which allows the transport of $\phi$ to be discretized through a flux-difference update.

\textit{Discrete update in lattice units.}
At a fluid node $(i,j,k)$, the discrete update is written as
\begin{equation}
\phi^{n+1}_{ijk}
=
\phi^{n}_{ijk}
-
(\nabla_h\cdot \mathbf{F}_a)_{ijk}
+
D
(\nabla_h^2\phi)_{ijk}
+
S^{\mathrm{comp}}_{ijk},
\label{eq:phi_update_muscl}
\end{equation}
where $\nabla_h$ denotes the discrete Cartesian operators and $S^{\mathrm{comp}}$ is the discrete form of the compressive term
\[
-\kappa\nabla\cdot
\left[
\phi(1-\phi)\mathbf{n}
\right].
\]
A double-buffering strategy is used for the phase field, and the update is applied only to fluid nodes, while solid nodes are excluded from the evolution step.

\subsubsection*{3D MUSCL--TVD minmod conservative advection for $\phi$}

\textit{Face fluxes and discrete divergence.}
Assuming unit lattice spacing, $\Delta x=\Delta y=\Delta z=1$, the discrete divergence of the advective flux is computed as
\begin{equation}
(\nabla_h\cdot\mathbf{F}_a)_{ijk}
=
\left(
F^x_{i+\frac12,j,k}
-
F^x_{i-\frac12,j,k}
\right)
+
\left(
F^y_{i,j+\frac12,k}
-
F^y_{i,j-\frac12,k}
\right)
+
\left(
F^z_{i,j,k+\frac12}
-
F^z_{i,j,k-\frac12}
\right),
\label{eq:divF}
\end{equation}
where the face fluxes are
\begin{equation}
F^x_{i+\frac12,j,k}
=
u_{i+\frac12,j,k}
\phi^{\mathrm{up}}_{i+\frac12,j,k},
\qquad
F^y_{i,j+\frac12,k}
=
v_{i,j+\frac12,k}
\phi^{\mathrm{up}}_{i,j+\frac12,k},
\qquad
F^z_{i,j,k+\frac12}
=
w_{i,j,k+\frac12}
\phi^{\mathrm{up}}_{i,j,k+\frac12}.
\label{eq:face_flux_def}
\end{equation}
The face-centered velocities are obtained by arithmetic interpolation of neighboring cell-centered values; for example,
\[
u_{i+\frac12,j,k}
=
\frac{1}{2}
\left(
u_{i,j,k}
+
u_{i+1,j,k}
\right),
\]
with analogous expressions for $v$ and $w$.

\textit{MUSCL reconstruction with minmod limiter.}
The upwind value $\phi^{\mathrm{up}}$ at each face is computed through a MUSCL piecewise-linear reconstruction limited by the minmod function, which ensures the TVD property. Along the $x$ direction, the limited slope at cell $(i,j,k)$ is
\begin{equation}
s^x_{i,j,k}
=
\mathrm{minmod}
\left(
\phi_{i,j,k}-\phi_{i-1,j,k},
\,
\phi_{i+1,j,k}-\phi_{i,j,k}
\right),
\label{eq:slope_minmod_x}
\end{equation}
with
\begin{equation}
\mathrm{minmod}(a,b)
=
\begin{cases}
0, & ab\le 0,\\
\mathrm{sign}(a)\min(|a|,|b|), & ab>0.
\end{cases}
\label{eq:minmod}
\end{equation}
The left and right reconstructed states at the face $i+\tfrac12$ are
\begin{equation}
\phi^{L}_{i+\frac12,j,k}
=
\phi_{i,j,k}
+
\frac{1}{2}s^x_{i,j,k},
\qquad
\phi^{R}_{i+\frac12,j,k}
=
\phi_{i+1,j,k}
-
\frac{1}{2}s^x_{i+1,j,k},
\label{eq:recon_lr_x}
\end{equation}
and the upwind state is selected as
\begin{equation}
\phi^{\mathrm{up}}_{i+\frac12,j,k}
=
\begin{cases}
\phi^{L}_{i+\frac12,j,k}, & u_{i+\frac12,j,k}\ge 0,\\
\phi^{R}_{i+\frac12,j,k}, & u_{i+\frac12,j,k}< 0.
\end{cases}
\label{eq:upwind_select_x}
\end{equation}
The same reconstruction and upwind selection are applied in the $y$ and $z$ directions by replacing $(i,u)$ with $(j,v)$ and $(k,w)$, respectively. The resulting scheme is second-order accurate in smooth regions, while reverting locally to first order near sharp gradients, thereby suppressing spurious oscillations at the interface.

\textit{Stencil requirements and boundary handling.}
The computation of the minmod slopes at neighboring cells requires access to phase-field values up to two grid points away in each coordinate direction. Two buffer layers are therefore used in each direction. In periodic domains, these layers are filled by wrap-around; otherwise, they are populated consistently with the imposed boundary conditions.

\subsubsection*{Discrete compressive term and isotropic divergence}

The compressive contribution in Eq.~\eqref{eq:cace} is written as the divergence of the compressive flux
\[
\mathbf{F}_c
=
\phi(1-\phi)\mathbf{n}.
\]
Its discrete divergence is evaluated using an isotropic D3Q27-based stencil that combines axis, face-diagonal, and body-diagonal contributions with weights $(p_1,p_2,p_3)$. Denoting by
\[
\mathbf{a}=(a_x,a_y,a_z)
\]
the discrete compressive vector field, or equivalently the discrete representation of $\mathbf{F}_c$, the source term is computed as
\begin{equation}
S^{\mathrm{comp}}_{ijk}
=
-\kappa
(\nabla_h\cdot\mathbf{a})_{ijk},
\label{eq:source_sharp}
\end{equation}
where $\kappa$ is the sharpening coefficient. The use of an isotropic stencil reduces grid-orientation errors and yields a compressive contribution consistent with the discrete operators used in the LB solver.

\subsection{Implementation of the contact angle in the conservative Allen-Cahn  equation model} \label{sec:wetting_bc}

The multiphase solver advances the phase-field variable
$\phi\in[0,1]$ through a conservative Allen--Cahn equation. Wetting at solid
boundaries is imposed by assigning ghost values of $\phi$ inside the solid
region. These ghost values enter the finite-difference stencils of the
Allen--Cahn operator at wall-adjacent fluid nodes and enforce the prescribed
contact angle in a purely geometric manner.

Let $\Omega_f$ and $\Omega_s$ denote the fluid and solid regions,
respectively, and let $\mathbf{x}_s\in\Omega_s$ be a solid node adjacent to the
fluid. The local geometry of the wall is reconstructed from the set of lattice
links connecting $\mathbf{x}_s$ to neighboring fluid nodes. Let
$\mathbf{c}_q$ be the discrete lattice vectors and define
\[
\mathcal{N}_f(\mathbf{x}_s)
=
\left\{
q:\mathbf{x}_s+\mathbf{c}_q\in\Omega_f
\right\}.
\]
An unnormalized local wall normal is obtained as
\begin{equation}
\widetilde{\mathbf{n}}_w(\mathbf{x}_s)
=
\sum_{q\in\mathcal{N}_f(\mathbf{x}_s)}
w_q\,(-\mathbf{c}_q),
\label{eq:local_wall_normal}
\end{equation}
where the weights depend on the lattice-link length,
\begin{equation}
w_q=
\begin{cases}
1, & |\mathbf{c}_q|^2=1,\\[1mm]
1/2, & |\mathbf{c}_q|^2=2,\\[1mm]
1/3, & |\mathbf{c}_q|^2=3.
\end{cases}
\label{eq:wall_normal_weights}
\end{equation}
The unit wall normal is then
\begin{equation}
\mathbf{n}_w
=
\frac{\widetilde{\mathbf{n}}_w}
{\left|\widetilde{\mathbf{n}}_w\right|}.
\label{eq:unit_wall_normal}
\end{equation}
With this convention, $\mathbf{n}_w$ points locally from the neighboring fluid
region towards the solid node. If
$\left|\widetilde{\mathbf{n}}_w\right|$ is smaller than a prescribed tolerance,
the node is skipped.

A donor fluid node $\mathbf{x}_f$ is then selected among the neighboring
fluid nodes. The donor is chosen as the fluid node most closely aligned with
the reconstructed wall normal. For each admissible lattice link, the alignment
score is
\begin{equation}
S_q
=
\frac{(-\mathbf{c}_q)\cdot\mathbf{n}_w}
{|\mathbf{c}_q|}.
\label{eq:donor_alignment_score}
\end{equation}
The selected donor corresponds to
\begin{equation}
q^\star
=
\operatorname*{arg\,max}_{q\in\mathcal{D}(\mathbf{x}_s)}
S_q,
\label{eq:donor_selection}
\end{equation}
where $\mathcal{D}(\mathbf{x}_s)$ is the set of admissible donor links. In
practice, wall-adjacent fluid nodes are considered first; if no such node is
available, the search is extended to all neighboring fluid nodes. The
donor-to-solid displacement vector is therefore
\begin{equation}
\mathbf{r}_{fs}
=
\mathbf{x}_s-\mathbf{x}_f
=
-\mathbf{c}_{q^\star}.
\label{eq:donor_to_solid}
\end{equation}

At the donor node, the phase-field value is denoted by
\[
\phi_f=\phi(\mathbf{x}_f),
\]
and the local gradient is reconstructed as
\[
\nabla\phi_f = \nabla\phi(\mathbf{x}_f).
\]
The component of the gradient tangent to the reconstructed wall is obtained by
removing the wall-normal projection:
\begin{equation}
\nabla_{\parallel}\phi_f
=
\nabla\phi_f
-
\left(
\nabla\phi_f\cdot\mathbf{n}_w
\right)\mathbf{n}_w .
\label{eq:tangential_gradient_contact_angle}
\end{equation}
The prescribed contact angle is imposed by requiring the corrected gradient to
form the desired angle with the wall normal. The angle used internally is
\begin{equation}
\theta_c = \pi-\theta,
\label{eq:internal_contact_angle}
\end{equation}
where $\theta$ is the prescribed physical contact angle. The
corresponding wall-normal derivative is
\begin{equation}
\partial_{n_w}\phi
=
-
\left|
\nabla_{\parallel}\phi_f
\right|
\cot\theta_c .
\label{eq:normal_derivative_contact_angle}
\end{equation}
Equivalently, the corrected gradient satisfying the geometric wetting
condition is
\begin{equation}
\nabla\phi_c
=
\nabla_{\parallel}\phi_f
+
\left(
\partial_{n_w}\phi
\right)
\mathbf{n}_w .
\label{eq:corrected_contact_angle_gradient}
\end{equation}
For limiting values of the contact angle, the cotangent is regularized by
replacing it with a large finite value when $|\sin\theta_c|$ falls below a
small numerical tolerance.

The ghost value inside the solid node is finally obtained by a first-order
extrapolation from the donor fluid node along the selected donor-to-solid
direction:
\begin{equation}
\phi_g
=
\phi_f
+
\lambda_s\,
\nabla\phi_c\cdot\mathbf{r}_{fs}.
\label{eq:ghost_contact_angle_extrapolation}
\end{equation}
The coefficient
\begin{equation}
\lambda_s
=
\sqrt{4\phi_f(1-\phi_f)},
\qquad
\phi_f=\phi(\mathbf{x}_f),
\label{eq:contact_angle_localization}
\end{equation}

acts as an interfacial localization factor. It is equal to one at the center
of the diffuse interface, $\phi_f=1/2$, and vanishes in the bulk phases,
$\phi_f=0$ or $\phi_f=1$. The wetting correction is therefore concentrated in
the contact-line region and does not modify ghost values far from the
interface.

The resulting ghost-node update is summarized in the following pseudo-code:

\begin{algorithmic}[1]

\ForAll{solid nodes $\mathbf{x}_s\in\Omega_s$ adjacent to the fluid}

  \State Identify the neighboring fluid links
  \Statex \hspace{\algorithmicindent}
  $\mathcal{N}_f(\mathbf{x}_s)
  =
  \{q:\mathbf{x}_s+\mathbf{c}_q\in\Omega_f,\ \mathbf{c}_q\neq\mathbf{0}\}$

  \If{$\mathcal{N}_f(\mathbf{x}_s)=\emptyset$}
     \State skip the node
  \EndIf

  \State Reconstruct the unnormalized wall normal
  \Statex \hspace{\algorithmicindent}
  $\widetilde{\mathbf{n}}_w
  =
  \sum_{q\in\mathcal{N}_f(\mathbf{x}_s)}
  w_q(-\mathbf{c}_q)$

  \State Normalize it as
  \Statex \hspace{\algorithmicindent}
  $\mathbf{n}_w
  =
  \widetilde{\mathbf{n}}_w/
  |\widetilde{\mathbf{n}}_w|$

  \State For each admissible donor link, compute the alignment score
  \Statex \hspace{\algorithmicindent}
  $S_q =
  \big[(-\mathbf{c}_q)\cdot\mathbf{n}_w\big]/|\mathbf{c}_q|$

  \State Select the donor link
  \Statex \hspace{\algorithmicindent}
  $q^\star =
  \operatorname*{arg\,max}_{q\in\mathcal{D}(\mathbf{x}_s)} S_q$

  \State Define the donor fluid node and donor-to-solid displacement
  \Statex \hspace{\algorithmicindent}
  $\mathbf{x}_f=\mathbf{x}_s+\mathbf{c}_{q^\star}$,
  \qquad
  $\mathbf{r}_{fs}=\mathbf{x}_s-\mathbf{x}_f=-\mathbf{c}_{q^\star}$

  \State Evaluate the phase-field value and gradient at the donor node
  \Statex \hspace{\algorithmicindent}
  $\phi_f=\phi(\mathbf{x}_f)$,
  \qquad
  $\nabla\phi_f=\nabla\phi(\mathbf{x}_f)$

  \State Remove the wall-normal projection
  \Statex \hspace{\algorithmicindent}
  $\nabla_{\parallel}\phi_f
  =
  \nabla\phi_f
  -
  (\nabla\phi_f\cdot\mathbf{n}_w)\mathbf{n}_w$

  \State Convert the prescribed contact angle according to the normal convention
  \Statex \hspace{\algorithmicindent}
  $\theta_c=\pi-\theta$

  \State Impose the geometric wetting condition
  \Statex \hspace{\algorithmicindent}
  $\partial_{n_w}\phi
  =
  -
  |\nabla_{\parallel}\phi_f|\cot\theta_c$

  \State Construct the corrected gradient
  \Statex \hspace{\algorithmicindent}
  $\nabla\phi_c
  =
  \nabla_{\parallel}\phi_f
  +
  (\partial_{n_w}\phi)\mathbf{n}_w$

  \State Compute the interfacial localization factor
  \Statex \hspace{\algorithmicindent}
  $\lambda_s=
  \sqrt{4\phi_f(1-\phi_f)}$,
  \qquad
  $\phi_f=\phi(\mathbf{x}_f)$

  \State Extrapolate the ghost value
  \Statex \hspace{\algorithmicindent}
  $\phi_g
  =
  \phi_f
  +
  \lambda_s
  \nabla\phi_c\cdot\mathbf{r}_{fs}$
  \State Assign the ghost value to the solid node
\Statex \hspace{\algorithmicindent}
$\phi(\mathbf{x}_s)\leftarrow\phi_g$

\EndFor

\end{algorithmic}
At the boundary level, each solid node reads only neighboring fluid data and
writes exclusively to its own ghost value. The wetting update is therefore
local and race-free, and is directly compatible with the thread-safe GPU
implementation of the hybrid lattice Boltzmann--Allen--Cahn solver. The global
mass-control procedure described in the next section is separate from this
boundary update and requires domain-wide reductions of scalar quantities.

\subsection{Mass conservation in Allen--Cahn interface tracking in the presence of wetting boundaries}
\label{sec:mass_conservation_wetting}

A conservative Allen--Cahn formulation preserves the total amount of each phase provided that the discrete phase-field flux satisfies a no-flux condition at solid boundaries. This property can be altered in the presence of wetting boundary conditions. In particular, for neutral wetting, corresponding to $\theta=90^\circ$, the imposed wall-normal derivative of the phase field vanishes. For non-neutral wetting, instead, the geometric contact-angle condition requires a non-zero wall-normal component of $\nabla\phi$ at the solid surface. In a diffuse-interface formulation, this produces a finite phase-field flux across the numerical boundary stencil. As a consequence, even if the bulk Allen--Cahn equation is written in conservative form, the combined action of the finite-difference stencil and the ghost-node wetting condition may lead to a small but systematic gain or loss of the total phase-field mass.

Here, by phase-field mass we denote the discrete liquid volume,
\begin{equation}
M_\phi(t)
=
\sum_{\mathbf{x}\in\Omega_f}
\phi(\mathbf{x},t)\,\Delta V,
\label{eq:phase_field_mass}
\end{equation}
where the sum is restricted to fluid nodes and $\Delta V$ is the lattice-cell volume. In lattice units, $\Delta V=1$. To remove the boundary-induced mass drift, we enforce a global conservation constraint at each time step by adding a corrective source or sink term to the Allen--Cahn update. This correction can be interpreted as a discrete Lagrange multiplier enforcing conservation of $M_\phi$.

Let $M_\phi^n$ be the phase-field mass at the beginning of the time step. After applying the wetting boundary condition and advancing the conservative Allen--Cahn equation, a tentative field $\phi^\star$ is obtained. Its corresponding mass is
\begin{equation}
M_\phi^\star
=
\sum_{\mathbf{x}\in\Omega_f}
\phi^\star(\mathbf{x})\,\Delta V .
\label{eq:tentative_phase_field_mass}
\end{equation}
The mass defect generated during the time step is then
\begin{equation}
\Delta M_\phi
=
M_\phi^n
-
M_\phi^\star .
\label{eq:mass_defect}
\end{equation}
A positive value of $\Delta M_\phi$ indicates that phase-field mass has been lost and must be re-injected, whereas a negative value indicates a mass excess that must be removed.

The correction is distributed only over the diffuse-interface region, rather than uniformly over the whole domain. This avoids modifying the bulk phases and preserves the sharpness of the interface. We define an interfacial weight
\begin{equation}
\chi(\mathbf{x})
=
\begin{cases}
\phi^\star(\mathbf{x})
\left[
1-\phi^\star(\mathbf{x})
\right],
&
\epsilon_\phi < \phi^\star(\mathbf{x}) < 1-\epsilon_\phi,
\\[1mm]
0,
&
\text{otherwise},
\end{cases}
\label{eq:interface_weight_mass_correction}
\end{equation}
where $\epsilon_\phi$ is a small numerical threshold used to identify the diffuse-interface band. The normalization factor is
\begin{equation}
\mathcal{W}
=
\sum_{\mathbf{x}\in\Omega_f}
\chi(\mathbf{x})\,\Delta V .
\label{eq:mass_correction_normalization}
\end{equation}
Provided that $\mathcal{W}>0$, the corrected phase field is obtained as
\begin{equation}
\phi^{n+1}(\mathbf{x})
=
\phi^\star(\mathbf{x})
+
\frac{\Delta M_\phi}{\mathcal{W}}\,
\chi(\mathbf{x}) .
\label{eq:mass_corrected_phi}
\end{equation}
Equivalently, this operation corresponds to adding to the conservative Allen--Cahn equation a source term of the form
\begin{equation}
S_M(\mathbf{x},t)
=
\lambda_M(t)\,\chi(\mathbf{x}),
\label{eq:mass_source_term}
\end{equation}
where the scalar multiplier $\lambda_M$ is chosen so that the integral constraint on $M_\phi$ is exactly satisfied. In discrete form,
\begin{equation}
\lambda_M
=
\frac{\Delta M_\phi}
{\Delta t\,\mathcal{W}} .
\label{eq:lagrange_multiplier_mass}
\end{equation}
In lattice units, $\Delta t=1$.

Since the correction is localized by $\chi$, it vanishes in the bulk regions where $\phi\simeq0$ or $\phi\simeq1$ and acts only where the diffuse interface is present. The procedure therefore compensates the net boundary-induced mass loss or gain without altering pure phases away from the interface. In practice, the correction is small at each time step but prevents the accumulation of systematic mass errors over long simulations, which is especially important for wetting-controlled dynamics in confined or porous geometries, where contact lines may remain in contact with solid boundaries for long times.

\section{Results} \label{sec:results}

\subsection{Static contact-angle test}
\label{sec:static_contact_angle}

The accuracy of the geometric wetting boundary condition was first assessed
through a series of static contact-angle tests. The benchmark consists of a
liquid droplet resting on a flat solid wall and surrounded by the second fluid.
The droplet is initialized as a spherical cap with an apparent contact angle of
approximately \(90^\circ\) and is subsequently allowed to relax toward the
equilibrium angle prescribed at the wall. This test therefore verifies whether
the ghost-node wetting condition drives the interface toward the target
macroscopic contact angle without introducing significant distortions of the
equilibrium droplet shape.

All simulations were performed in a cubic domain of size
\[
n_x\times n_y\times n_z=256^3
\]
lattice nodes. Unless otherwise stated, all physical and numerical parameters
are expressed in lattice units. The initial droplet radius was set to $R_0=40$, 
with its center located at $(x_c,y_c,z_c)=\left(\frac{n_x}{2},\frac{n_y}{2},2\right).
$
The bottom boundary was treated as a stationary solid wall, and wettability was
imposed using the ghost-node construction described in
Section~\ref{sec:methods}. The two fluids were assigned equal densities and
equal kinematic viscosities,
\[
\rho_1=\rho_2=1,
\qquad
\nu_1=\nu_2=0.04,
\]
while the surface tension was set to
\[
\sigma=0.02.
\]

For each prescribed contact angle, the simulation was advanced until the
droplet reached a stationary equilibrium configuration. The corresponding
numerical contact angle was then evaluated a posteriori from the equilibrium
phase-field distribution.

\begin{figure}[t]
    \centering
    \includegraphics[width=0.5\linewidth]{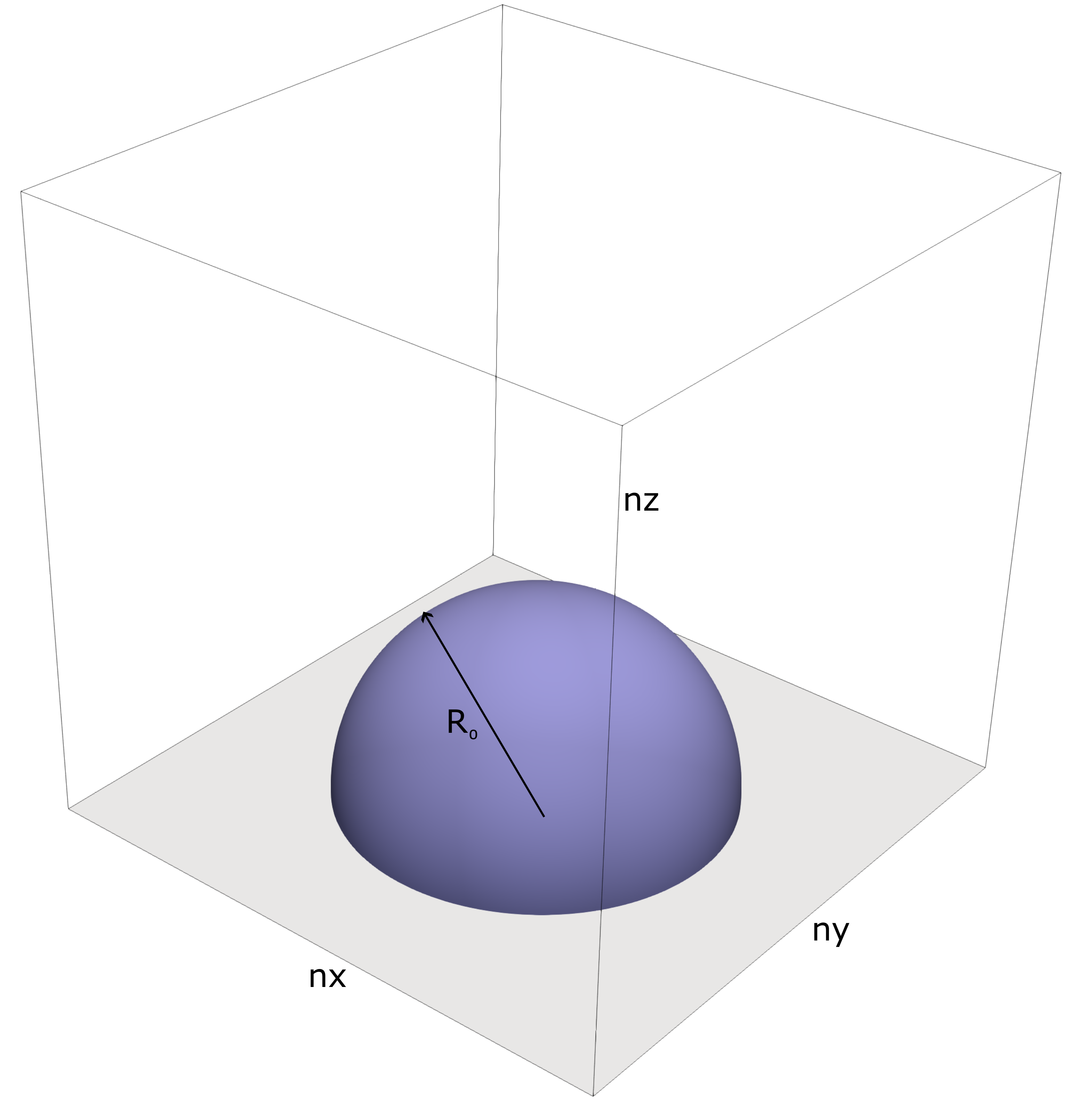}
    \caption{Schematic of the static contact-angle benchmark. A droplet of
    initial radius \(R_0\) is placed on the bottom solid wall of a cubic domain
    of size \(n_x\times n_y\times n_z\) and allowed to relax toward the
    prescribed equilibrium contact angle.}
    \label{fig:sketch_case1}
\end{figure}

\begin{figure}[t]
    \centering
    \includegraphics[width=0.55\linewidth]{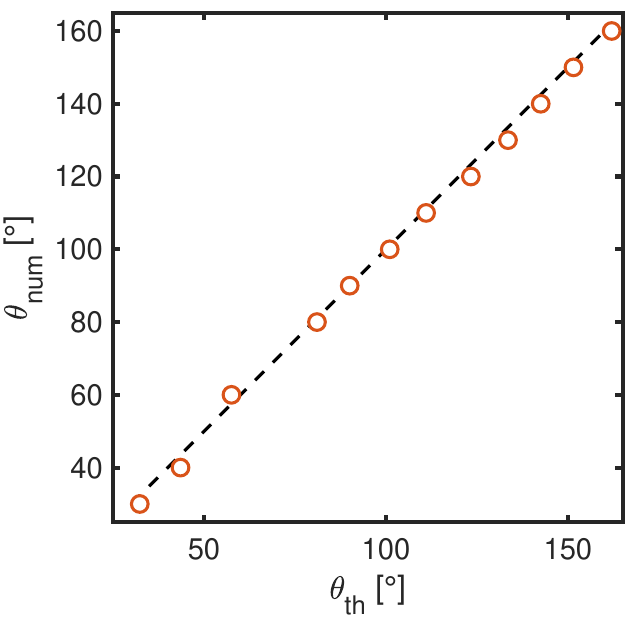}
    \caption{Static contact-angle validation. The measured numerical contact
    angle \(\theta_{\rm num}\) is compared with the prescribed theoretical
    contact angle \(\theta_{\rm th}\). Open circles denote the numerical
    measurements, while the dashed line represents the ideal identity relation
    \(\theta_{\rm num}=\theta_{\rm th}\). The close agreement over the range
    \(30^\circ\leq\theta_{\rm th}\leq160^\circ\) demonstrates the accuracy of
    the geometric wetting boundary condition from strongly hydrophilic to
    strongly hydrophobic conditions.}
    \label{fig:static_contact_angle}
\end{figure}

Figure~\ref{fig:static_contact_angle} compares the measured numerical contact
angle with the prescribed theoretical value. The dashed line represents the
ideal identity relation
\[
\theta_{\rm num}=\theta_{\rm th},
\]
while the open symbols denote the numerical measurements. The results remain
close to the identity line over the entire range investigated,
$
30^\circ\leq\theta_{\rm th}\leq160^\circ,
$
covering strongly hydrophilic, neutral-wetting, and strongly hydrophobic
conditions. Only small deviations of a few degrees are observed, including at
the limiting contact angles. This agreement demonstrates that the proposed
geometric wetting condition accurately recovers the prescribed macroscopic
contact angle over a wide range of wettabilities.

The contact angle was measured from a two-dimensional vertical section of the
final droplet configuration. The phase-field distribution was extracted on the
central \(xz\) plane, and the diffuse interface was identified from the points
satisfying
\[
\left|\phi(x,z)-\phi_0\right|<\epsilon_{fit},
\qquad
\phi_0=\frac{1}{2},
\]
where \(\epsilon_{fit}\) is a prescribed tolerance. 

The resulting \(N_{\rm points}\) interface coordinates were fitted with a
circle,
\[
(x-x_0)^2+(z-z_0)^2=R^2,
\]
by determining \((x_0,z_0,R)\) through a least-squares minimization of the
algebraic residual
\begin{equation}
\mathcal{E}
=
\sum_{i=1}^{N_{\rm points}}
\left[
(x_i-x_0)^2+(z_i-z_0)^2-R^2
\right]^2.
\label{eq:circle_fit_error}
\end{equation}
Here, \(R\) denotes the radius of the fitted circle and should not be confused
with the initial droplet radius \(R_0\).

The apparent contact angle was then reconstructed geometrically from the
intersection between the fitted circle and the wall. Denoting by \(z_w\) the
wall position used in the measurement, the numerical contact angle was
calculated as
\begin{equation}
\theta_{\rm num}
=
\pi-
\cos^{-1}
\left(
\frac{z_0-z_w}{R}
\right).
\label{eq:measured_contact_angle}
\end{equation}
The argument of the inverse cosine was clipped to the interval \([-1,1]\) to
avoid numerical round-off errors. This procedure provides a robust estimate of
the apparent macroscopic contact angle from the relaxed diffuse-interface
configuration.

\subsection{Short-time dynamics of a spreading droplet on a surface}
\label{sec:short_time_spreading}

As a second validation of the wetting boundary condition, we consider the
short-time spreading dynamics of a liquid droplet initially brought into contact
with a solid surface. This benchmark is inspired by the experiments and scaling
analysis of Bird, Mandre and Stone~\cite{bird2008short}, who showed that the
early spreading of partially wetting droplets is governed primarily by an
inertial--capillary balance. In particular, the spreading exponent is not
universal, but decreases systematically as the equilibrium contact angle is
increased.

The numerical simulations are performed in a cubic domain of size
\[
n_x \times n_y \times n_z = 512^3
\]
lattice nodes. A spherical droplet of initial radius
$
R_0 = 80
$
lattice units is initialized in contact with the bottom solid wall and is then
allowed to spread freely. No external body force is applied, so that the
transient dynamics is driven solely by capillarity and by the wetting condition
imposed at the wall.

The two fluids have equal densities and equal kinematic viscosities,
\[
\rho_1=\rho_2=1,
\qquad
\nu_1=\nu_2=0.05,
\]
while the surface tension is set to
\[
\sigma=0.02.
\]
The corresponding Ohnesorge number, evaluated using the initial droplet radius,
is
\[
Oh_{R_0}
=
\frac{\rho\nu}{\sqrt{\rho\sigma R_0}}
=
\nu\sqrt{\frac{\rho}{\sigma R_0}}
\simeq 3.95\times10^{-2},
\]
indicating that viscous effects remain secondary during the earliest stage of
the spreading process.

Seven equilibrium contact angles are investigated,
\[
\theta_{\rm eq}
=
40^\circ,\;55^\circ,\;70^\circ,\;85^\circ,\;100^\circ,\;115^\circ,\;130^\circ,
\]
thereby covering continuously the transition from hydrophilic to strongly
hydrophobic conditions.

\begin{figure}[t]
    \centering
    \includegraphics[width=0.5\linewidth]{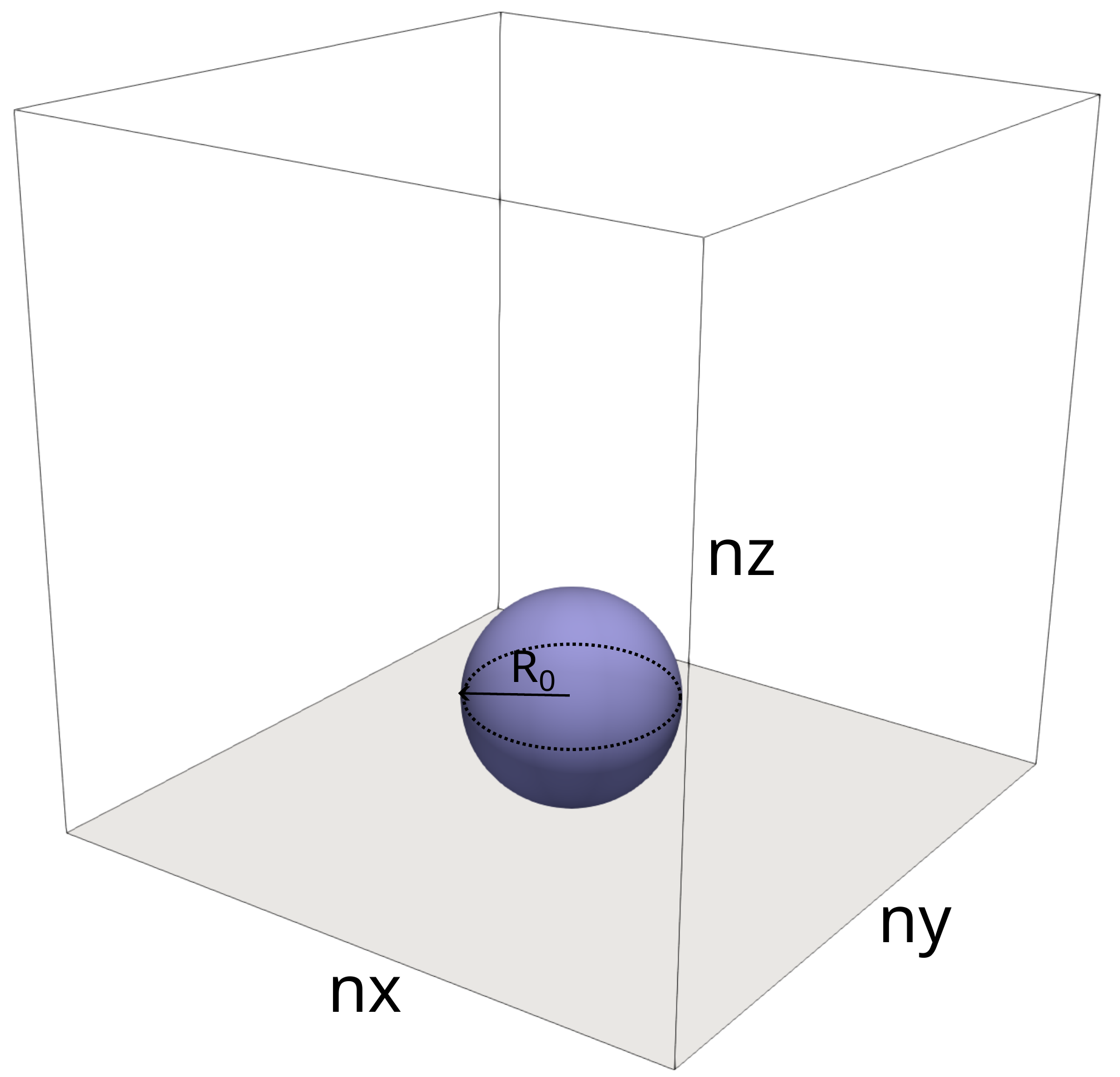}
    \caption{Schematic of the short-time spreading benchmark. A spherical
    droplet of initial radius \(R_0\) is placed in contact with the bottom solid
    wall and allowed to spread under the combined action of capillarity and the
    prescribed wetting condition. No external body force is applied.}
    \label{fig:sketch_case2}
\end{figure}

The spreading dynamics is quantified through the instantaneous wetted radius
\(r(t)\), normalized by the initial droplet radius \(R_0\). Time is made
dimensionless using the inertial--capillary time scale
\begin{equation}
\tau_i
=
\left(
\frac{\rho R_0^3}{\sigma}
\right)^{1/2},
\label{eq:inertial_time_spreading}
\end{equation}
and the dimensionless time is defined as
$
t^*=\frac{t}{\tau_i}.
$
The spreading curves are therefore reported in terms of
\[
\frac{r(t)}{R_0}
\qquad \text{versus} \qquad
t^*.
\]
This normalization isolates the inertial--capillary dynamics and permits a
direct comparison of the early spreading behavior across the different
wettability conditions.

\begin{figure}[t]
    \centering
    \includegraphics[scale=0.8]{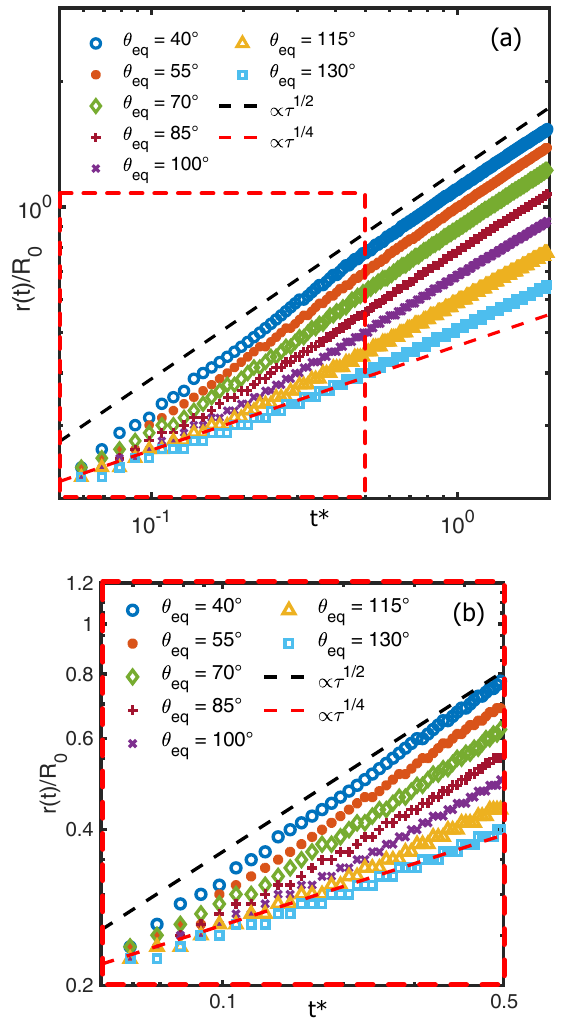}
    \caption{Short-time spreading dynamics for different equilibrium contact
angles. (a) Normalized wetted radius \(r(t)/R_0\) as a function of the
inertial--capillary dimensionless time \(t^*=t/\tau_i\), with
\(\tau_i=(\rho R_0^3/\sigma)^{1/2}\), for
\(\theta_{\rm eq}=40^\circ,55^\circ,70^\circ,85^\circ,100^\circ,
115^\circ\), and \(130^\circ\). The dashed lines indicate the reference
power laws \(r/R_0\propto t^{*1/2}\) and
\(r/R_0\propto t^{*1/4}\). (b) Enlargement of the early-time interval
highlighted by the dashed box in panel~(a).}
    \label{fig:short_time_spreading}
\end{figure}

Figure~\ref{fig:short_time_spreading} shows a systematic effect of
wettability on the contact-line dynamics. Hydrophilic droplets spread more
rapidly and attain a larger wetted radius at a given dimensionless time,
whereas increasing \(\theta_{\rm eq}\) progressively slows the spreading.

During the early inertial--capillary regime, the dynamics can be represented
by the effective power law
\begin{equation}
\frac{r(t)}{R_0}
\propto
\left(
\frac{t}{\tau_i}
\right)^{\alpha_{\theta}},
\label{eq:spreading_power_law}
\end{equation}
where the exponent \(\alpha_{\theta}\) depends on the prescribed equilibrium
contact angle. As shown more clearly in panel~(b), the numerical curves lie
between the reference scalings \(t^{*1/2}\) and \(t^{*1/4}\). The most
hydrophilic case, \(\theta_{\rm eq}=40^\circ\), approaches the classical
inertial scaling \(\alpha_{\theta}\simeq1/2\), whereas the most hydrophobic
cases, \(\theta_{\rm eq}=115^\circ\) and \(130^\circ\), approach
\(\alpha_{\theta}\simeq1/4\). Intermediate contact angles produce a continuous
ordering between these limiting behaviors.

The results are therefore consistent with the contact-angle-dependent
spreading dynamics reported by Bird, Mandre and Stone
\cite{bird2008short}, and show that the proposed wetting treatment captures
the transient response to surface wettability in addition to the prescribed
static contact angle.

\subsection{Maximal deformation of an impacting drop on a hydrophobic surface}
\label{sec:maximal_deformation}

As a further validation of the dynamic wetting behavior, we consider the
impact of a liquid droplet on a hydrophobic solid surface. This benchmark
follows the experiments and scaling analysis of Clanet et
al.~\cite{clanet2004maximal}, who showed that the maximum deformation of a
low-viscosity droplet impacting a non-wetting surface follows the
inertial--capillary scaling
\begin{equation}
\frac{D_{\max}}{D_0}\propto We^{1/4},
\label{eq:clanet_scaling}
\end{equation}
where \(D_0\) is the initial droplet diameter and \(D_{\max}\) is the maximum
diameter attained during spreading.

The simulations are performed in a cubic domain of size
$
n_x\times n_y\times n_z=512^3
$
lattice nodes. A spherical droplet of initial radius
$
R_0=80$ (
$
D_0=2R_0=160,
$)
is initialized with its lower interface in contact with the bottom solid wall.
The droplet is assigned a uniform initial velocity \(U_0\) directed normally
towards the wall. Gravity is not applied, so that the impact velocity can be
controlled directly and independently of the initial droplet position.

A strongly hydrophobic equilibrium contact angle,
$
\theta_{\rm eq}=130^\circ,
$
is imposed at the solid surface. The liquid and gas densities are set to
\[
\rho_{\ell}=10,
\qquad
\rho_g=0.01,
\]
corresponding to a liquid-to-gas density ratio
\[
\frac{\rho_{\ell}}{\rho_g}=1000.
\]
The two phases have equal kinematic viscosities,
\[
\nu_{\ell}=\nu_g=0.05,
\]
and the surface tension is
\[
\sigma=0.02.
\]
The liquid Ohnesorge number, based on the initial droplet diameter, is therefore
\begin{equation}
Oh=
\frac{\mu_{\ell}}
{\sqrt{\rho_{\ell}\sigma D_0}}
=
\nu_{\ell}
\sqrt{\frac{\rho_{\ell}}{\sigma D_0}}
\simeq 8.8\times10^{-2},
\label{eq:oh_clanet}
\end{equation}
placing the simulations in a regime in which inertia and capillarity provide
the dominant contributions to the maximum deformation, although viscous
effects are not entirely negligible.
\begin{figure}
    \centering
    \includegraphics[width=1\linewidth]{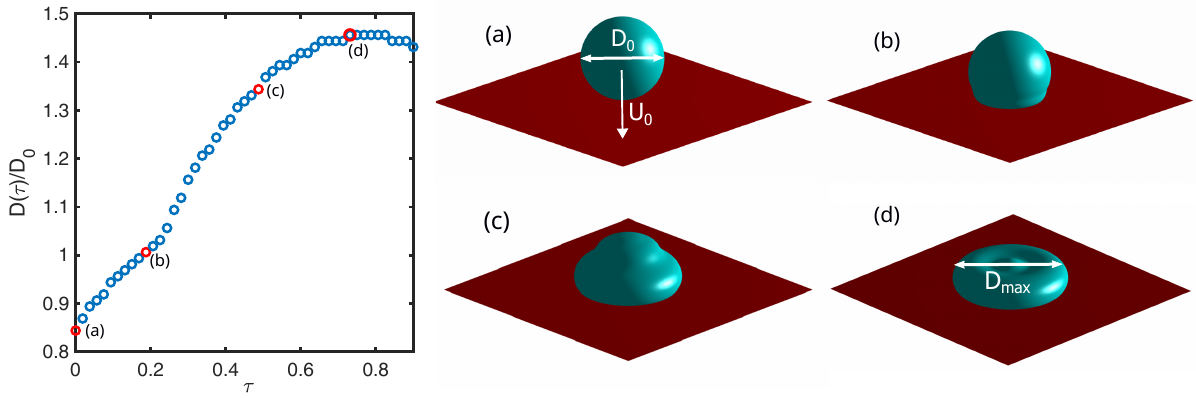}
    \caption{Representative stages of droplet impact on a hydrophobic surface:
    the left plot report the non dimensional spreading diameter versus the non dimensional time $\tau=t/(D_0/U_0)$
(a) initial configuration, with droplet diameter \(D_0\) and impact velocity
\(U_0\); (b,c) intermediate deformation during impact and spreading; and
(d) configuration at maximum lateral extension \(D_{\max}\).}
\label{fig:impact_sequence}
\end{figure}
The impact Weber number is varied by changing the initial velocity \(U_0\)
while keeping all other physical and geometrical parameters fixed:
\begin{equation}
We=
\frac{\rho_{\ell}U_0^2D_0}{\sigma}.
\label{eq:we_clanet}
\end{equation}
For each simulation, the instantaneous lateral extension of the droplet is
measured in the direction parallel to the wall. Its maximum value before the
onset of retraction defines \(D_{\max}\). The resulting maximum spreading
factor is
\begin{equation}
\beta_{\max}=
\frac{D_{\max}}{D_0}.
\label{eq:beta_max}
\end{equation}

A representative sequence of the impact and spreading dynamics for $We\simeq7$ is shown in
Fig.~\ref{fig:impact_sequence}.

\begin{figure}
    \centering
    \includegraphics[width=0.6\linewidth]{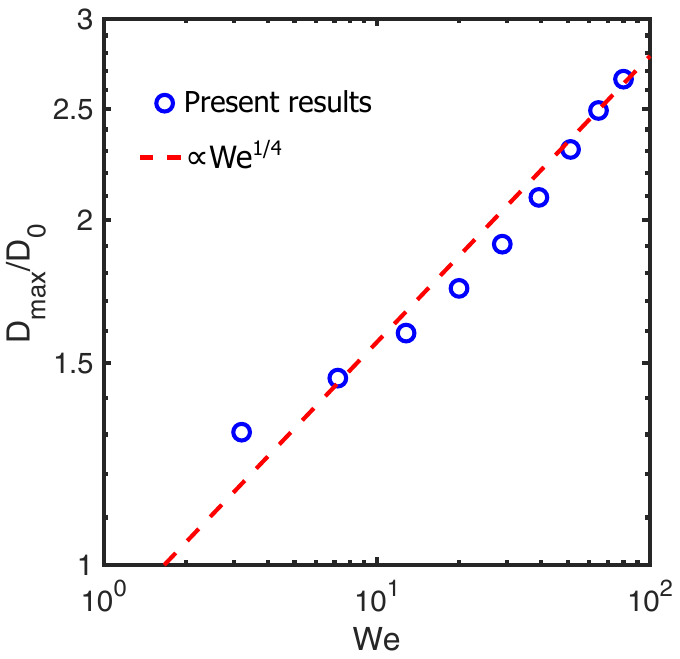}
    \caption{Maximum deformation of a droplet impacting a hydrophobic surface.
The maximum spreading factor \(D_{\max}/D_0\) is reported as a function of the
impact Weber number \(We=\rho_{\ell}U_0^2D_0/\sigma\). Symbols denote the
present numerical results for \(\theta_{\rm eq}=130^\circ\), while the dashed
line represents the inertial--capillary scaling
\(D_{\max}/D_0\propto We^{1/4}\) proposed by Clanet et
al.~\cite{clanet2004maximal}.}
\label{fig:clanet}
\end{figure}

Figure~\ref{fig:clanet} reports \(\beta_{\max}\) as a function of the Weber
number. Increasing \(We\) produces a progressively larger maximum spreading
diameter, since the initial kinetic energy and momentum of the droplet become
increasingly important relative to the capillary stresses opposing
deformation. Over the investigated range, the maximum spreading factor
increases from approximately \(1.15\) at the lowest Weber number to about
\(2.7\) at the largest one.
The numerical results closely follow the reference scaling
\[
\beta_{\max}\propto We^{1/4}
\]
over most of the investigated Weber-number range. A modest departure is
observed for the lowest-\(We\) case, for which the deformation induced by the
impact is comparable to the initial wetting-driven deformation and a
well-developed inertial spreading regime is not yet established. As the Weber
number is increased, the numerical points progressively approach the
\(We^{1/4}\) power law and remain approximately parallel to it over more than one
decade in \(We\).

The agreement with the scaling proposed by Clanet et
al.~\cite{clanet2004maximal} demonstrates that the present formulation
correctly captures the competition between impact inertia and capillary
restoration during the strong deformation of a droplet on a hydrophobic
surface. Together with the short-time spreading benchmark, this test shows
that the geometric wetting boundary condition remains accurate not only during
spontaneous contact-line motion, but also under finite-velocity impact, large
interfacial deformation and a liquid-to-gas density ratio of \(1000\).

\subsection{Drop motion through a sharp-edged orifice}

\begin{figure}[t]
    \centering

   \includegraphics[width=0.8\linewidth]{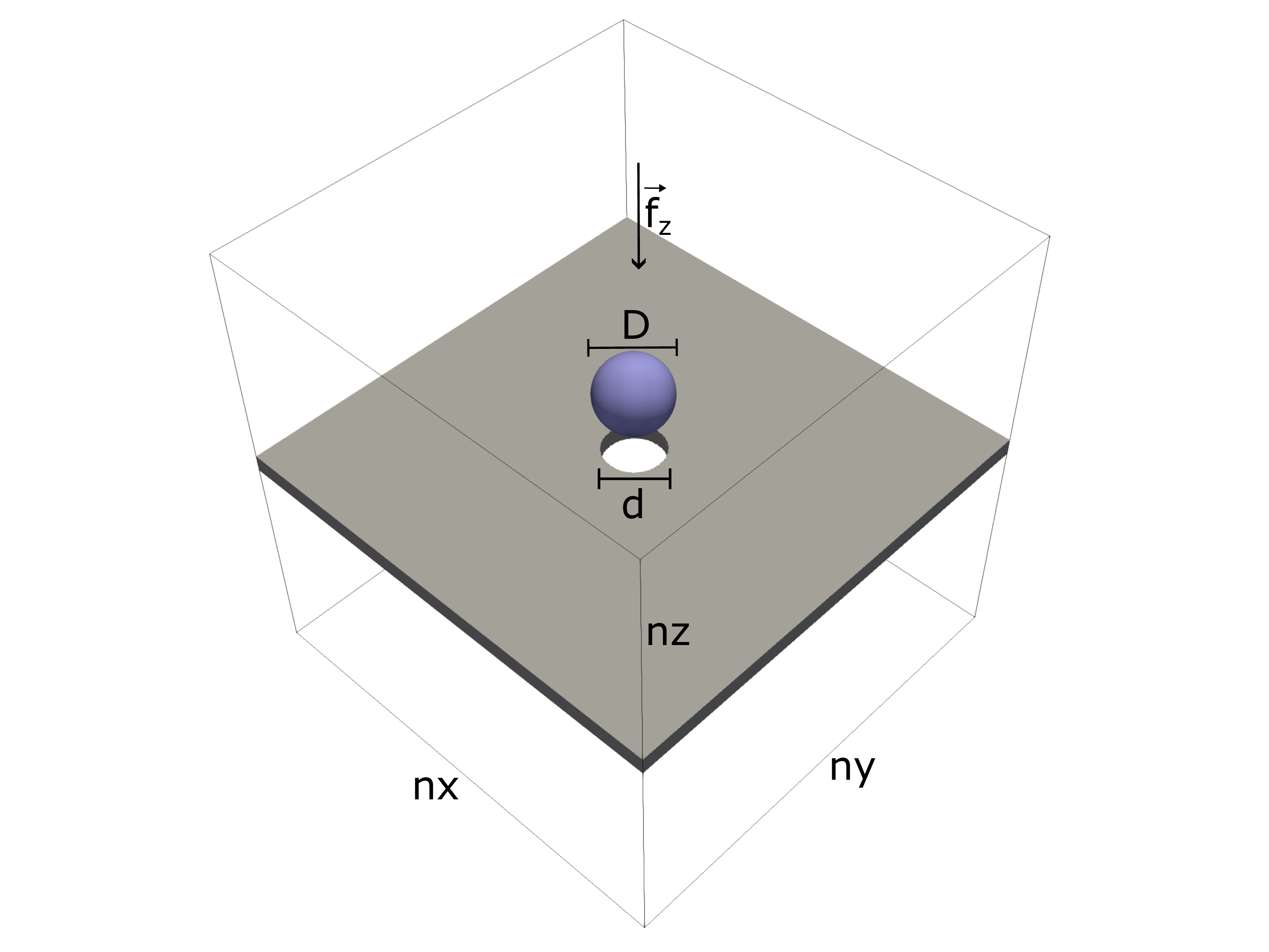}
   \caption{
    Schematic of the gravity-driven droplet-orifice benchmark. 
    A droplet of diameter $D$ is driven by the body force $f_z$ through a sharp-edged aperture of diameter $d$, with confinement set by the ratio $d/D$.
    }
    \label{fig:orifice_cases}
\end{figure}

\begin{figure}
    \centering
    \includegraphics[width=0.6\linewidth]{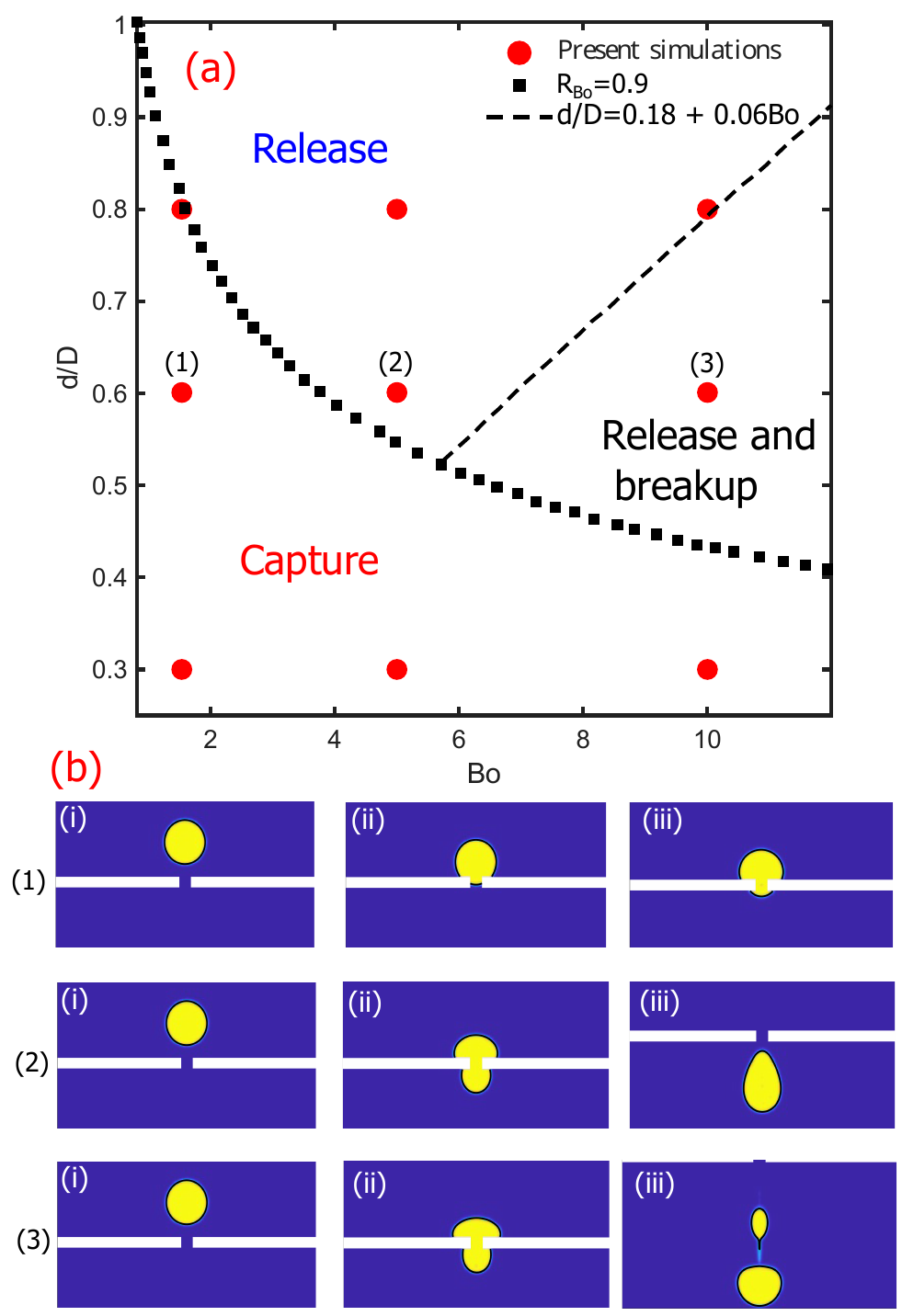}
\caption{
Drop motion through a sharp-edged orifice. (a) Regime map in the
\((Bo,d/D)\) plane. The reference curves indicate the capture--release and release--breakup
boundaries reported by Bordoloi and Longmire 
\cite{Bordoloi_Longmire_2014}. (b) Representative phase-field
snapshots at \(d/D=0.6\), showing capture, release, and release with
breakup as \(Bo\) is increased.
}
\label{fig:orifice}
\end{figure}

The fourth benchmark considers the motion of a deformable droplet through a sharp-edged circular orifice, following the experimental configuration investigated by Bordoloi and Longmire \cite{Bordoloi_Longmire_2014}. In that study, a liquid drop driven by gravity interacts with a confining opening whose diameter is smaller than the drop size. Depending on the balance between gravitational forcing, capillary resistance, confinement, and wetting, different outcomes are observed, including capture, release, and release accompanied by breakup. This configuration represents a stringent test for the numerical model, since in the sharp-edged case the interface comes into direct contact with the solid rim of the orifice, making contact-line dynamics, wetting, and possible pinning at the edge central to the subsequent evolution.

The overall behavior can be described in terms of a set of dimensionless parameters. The Bond number,
\[
Bo = \frac{\Delta \rho\, g\, D^2}{\sigma},
\]
the ratio $\frac{d}{D}$ where $d$ is the orifice diameter and $D$ is the initial droplet diameter, the Weber number,
\[
We = \frac{\rho_d U^2 D}{\sigma},
\]
and the Ohnesorge number,
\[
Oh = \frac{\mu_d}{\sqrt{\rho_d \sigma D}}.
\]

Further, we also report the modified Bond-number
parameter
\[
\mathcal{R}_{Bo}
=
\left[Bo\left(\frac{d}{D}\right)^3\right]^{1/2},
\]
defined as the ratio between the orifice capillary time scale and the drop
gravitational time scale \cite{Bordoloi_Longmire_2014}.

Here, $\Delta \rho$ denotes the density difference between the two phases, $g$ the acceleration driving the motion, $\sigma$ the surface tension, $\rho_d$ and $\mu_d$ the droplet density and dynamic viscosity, and $U$ a characteristic impact velocity. 

Consistently with the experimental observations, the global regimes are primarily organized in the $(Bo,d/D)$ plane, which determines whether the droplet remains captured or is able to pass through the orifice. Nevertheless, inertia remains relevant during the impact and early deformation stages, while viscous dissipation, although generally subdominant in the parameter range considered here, still affects the transient dynamics. The sharp-edged geometry further amplifies the role of wettability, since the contact line interacts with a geometrically singular edge where local pinning and depinning mechanisms can strongly influence the drop motion.

The orifice benchmark is carried out in a cubic domain of size $512^3$ lattice nodes. A single droplet is initialized on the symmetry axis of the orifice, with radius $R=40$ lattice units, corresponding to an initial droplet diameter $D=80$. The initial center of the droplet is placed at
\[
(x_0,y_0,z_0)=(nx/2,ny/2,nz/2 + D),
\]
i.e. two droplet radii above the center of the domain. The solid plate is located at mid-height, $z=256$, and has thickness
\[
h = \frac{2R}{4}=20
\]
lattice units. The orifice is centerd at $(x,y)=(nx/2,ny/2)$ and consists of a sharp-edged cylindrical opening through the plate. The orifice diameter is varied through the ratio \(d/D\), as shown in
Fig.~\ref{fig:orifice_cases}, while the edge of the opening forms a right
angle with the plate surface.

The diffuse-interface thickness is set to $\delta=6$ lattice units and the surface-tension parameter to $\sigma=0.025$. The two phases have densities $\rho_l=1.2$ and $\rho_g=0.9$, and kinematic viscosities $\nu_1=0.031$ and $\nu_2=0.24$, respectively. The prescribed equilibrium contact angle at the solid surface is fixed to $\theta=110^\circ$ in all simulations, consistently with the experimental configuration of Bordoloi and Longmire \cite{Bordoloi_Longmire_2014}. 

A constant body force is applied along the negative $z$ direction to drive the droplet toward the orifice. In the simulations reported here, the Bond number is varied by changing the magnitude of this vertical forcing, while all other physical parameters are kept fixed. 

Nine simulations were performed by varying \(Bo\) and \(d/D\).
Figure~\ref{fig:orifice} compares the resulting outcomes with the
experimental regime map of Bordoloi and Longmire
\cite{Bordoloi_Longmire_2014}. The numerical points fall within the
expected capture, release, and release-with-breakup regions. In
particular, droplets with
\(\mathcal{R}_{Bo}\lesssim0.9\) remain captured, whereas larger values
promote passage through the orifice. At sufficiently large \(Bo\), the
strong axial deformation generated during passage leads to necking and
breakup.

The snapshots in Fig.~\ref{fig:orifice}(b), obtained at \(d/D=0.6\),
illustrate this sequence. At low \(Bo\), capillary resistance and
contact-line pinning retain the droplet above the plate. Increasing
\(Bo\) first produces complete release without fragmentation and,
eventually, release accompanied by breakup.

\begin{figure}
    \centering
    \includegraphics[width=1.0\linewidth]{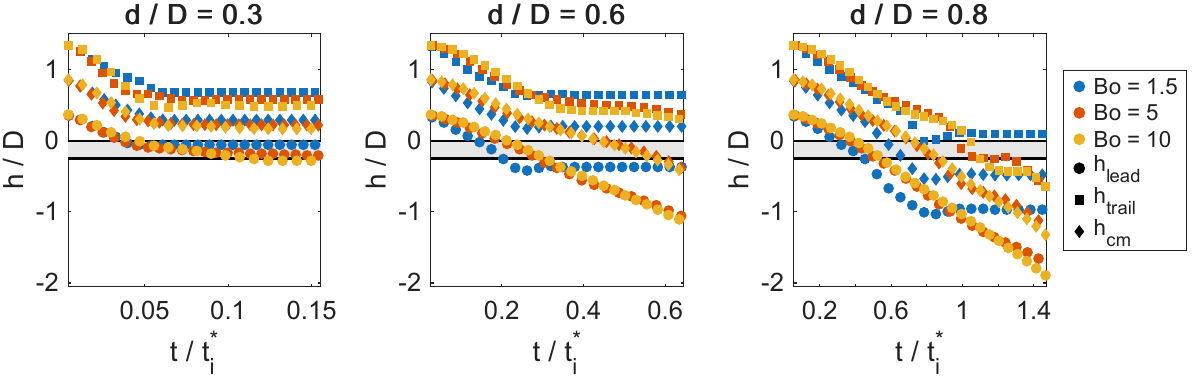}
\caption{Axial dynamics of a droplet crossing a sharp-edged orifice. Time evolution of
the leading edge $h_{\rm lead}$, trailing edge $h_{\rm trail}$, and center of mass
$h_{\rm cm}$, normalized by the droplet diameter $D$, for different confinement
ratios $d/D$ and Bond numbers $Bo$. The characteristic time is defined as $t_i^*=D^3/(Ud^2)$. The horizontal line $h/D=0$ marks the
orifice plane. Capture occurs when the center of mass remains above the orifice,
even if the leading edge partially penetrates it. Complete release is associated
with the downward motion of both the center of mass and the trailing interface,
while strong axial stretching leads to breakup.
}
\label{fig:lead_tral_cent}
\end{figure}

To characterize the passage dynamics beyond the final regime
classification, Fig.~\ref{fig:lead_tral_cent} reports the axial
positions of the leading edge, \(h_{\rm lead}\), trailing edge,
\(h_{\rm trail}\), and droplet center of mass, \(h_{\rm cm}\).
The positions are normalized by the initial droplet diameter \(D\),
with \(h/D=0\) corresponding to the orifice plane. Negative values
therefore indicate penetration below the plate.

The leading edge alone does not provide an unambiguous criterion for
release, since the droplet may partially enter the aperture while most
of its volume remains above the plate. Capture is identified when
\(h_{\rm cm}\) relaxes to a positive value, whereas complete release
requires a sustained downward motion of the center of mass together
with the passage of the trailing interface through the orifice.

For \(d/D=0.3\), the center of mass and trailing interface remain
above the plate over the investigated range of \(Bo\), indicating that
confinement, capillary resistance, and edge pinning prevent release.
At \(d/D=0.6\), the lowest-\(Bo\) case exhibits partial penetration,
with \(h_{\rm lead}<0\) but \(h_{\rm cm}>0\), whereas increasing the
forcing promotes passage of both the center of mass and trailing
interface. For \(d/D=0.8\), the weaker confinement produces deeper
penetration and stronger axial stretching; at sufficiently large
\(Bo\), the separation between the leading and trailing interfaces
culminates in necking and breakup.

These results show that the transition from capture to release is
controlled by the coupled evolution of penetration, center-of-mass
motion, and trailing-edge depinning, rather than by the first crossing
of the leading interface alone. The benchmark therefore tests the
wetting treatment under the simultaneous effects of gravity,
capillarity, confinement, edge pinning, large deformation, and
breakup.

\section{Conclusions}

We have developed and validated a geometric wetting boundary condition for a
conservative Allen--Cahn-based lattice Boltzmann framework. The prescribed
contact angle is enforced through ghost phase-field values constructed from a
wall normal reconstructed from the local fluid--solid occupancy and an explicit
donor-to-solid extrapolation. The formulation therefore does not require a
predefined Cartesian wall direction and can be applied directly to voxelized
solid geometries and sharp edges. The ghost-node update reads only neighboring
fluid data and is consequently local, race-free, and compatible with the
thread-safe implementation of the underlying lattice Boltzmann solver. A
separate interface-localized volume correction provides global control of the
phase-field mass and compensates for the small imbalance that may be introduced
by non-neutral wetting boundary conditions.

The method was assessed through four benchmarks addressing both equilibrium and
dynamic wetting. Static droplets recovered the prescribed contact angle over a
broad range of wettabilities. Short-time spreading reproduced the systematic
dependence of the inertial--capillary dynamics on the equilibrium contact angle,
with the numerical curves ranging between the limiting \(t^{1/2}\) and
\(t^{1/4}\) behaviors. Simulations of droplet impact on a hydrophobic surface
followed the classical \(We^{1/4}\) maximum-deformation scaling, including at a
liquid-to-gas density ratio of \(1000\). Finally, the gravity-driven passage of
a droplet through a sharp-edged orifice reproduced the experimentally observed
capture, release, and release-with-breakup regimes. The evolution of the
leading edge, trailing edge, and center of mass further showed that penetration
of the front interface alone does not imply complete release, which also
requires trailing-edge depinning and sustained passage of the droplet mass
through the aperture.

These results indicate that the proposed formulation provides an accurate and
computationally efficient treatment of wetting-controlled multiphase flows
involving contact-line motion, large interface deformation, confinement, and
interaction with sharp solid features. The present validation focuses mainly on
planar walls, complemented by the sharp-edged orifice configuration; a broader
assessment on curved and rough surfaces would be required to establish its
accuracy over fully general solid geometries. Nevertheless, the combination of
a geometrically defined ghost-node update, phase-field mass control, and
compatibility with high-performance lattice Boltzmann implementations makes the
method a promising basis for large-scale simulations in porous media,
microfluidic devices, and other geometrically confined multiphase systems.

\bibliography{apssamp}

\end{document}